# Prediction of two-dimensional nodal-line semimetal in a carbon nitride covalent network

Haiyuan Chen,[a,b] Shunhong Zhang,[c,b] Wei Jiang,[b] Chunxiao Zhang,[b] Heng Guo,[a] Zheng Liu,[c] Zhiming Wang,[a] Feng Liu,[b,d*] and Xiaobin Niu[a*]

[a] Institute of Fundamental and Frontier Sciences and School of Materials and Energy, University of Electronic Science and Technology of China, Chengdu 610054, P R China

[b] Department of Materials Science and Engineering, University of Utah, Salt Lake City, UT 84112, USA

[c] Institute for Advanced Study, Tsinghua Unviersity, Beijing 100084, PR China

[d] Collaborative Innovation Center of Quantum Matter, Beijing 100084, PR China

Corresponding author: xbniu@uestc.edu.cn, fliu@eng.utah.edu

**Abstract**

Carbon nitride compounds have emerged recently as a prominent member of 2D materials beyond graphene. The experimental realizations of 2D graphitic carbon nitride g-$C_3N_4$, nitrogenated holey grahpene $C_2N$, polyaniline $C_3N$ have shown their promising potential in energy and environmental applications. In this work, we predict a new type of carbon nitride network with a $C_9N_4$ stoichiometry from first principle calculations. Unlike common C-N compounds and covalent organic frameworks (COFs), which are typically insulating, surprisingly $C_9N_4$ is found to be a 2D nodal-line semimetal (NLSM). The nodal line in $C_9N_4$ forms a closed ring centered at $\Gamma$ point, which originates from the $p_z$ orbitals of both C and N. The linear crossing happens right at Fermi level contributed by two sets of dispersive Kagome and Dirac bands, which is robust due to negligible spin-orbital-coupling (SOC) in C and N. Besides, it is revealed that the formation of nodal ring is of accidental band degeneracy in nature induced by the chemical potential difference of C and N, as validated by a single orbital tight-binding model, rather than protected by crystal in-plane mirror symmetry or band topology. Interestingly, a new structure of nodal line, i.e., nodal-cylinder, is found in momentum space for AA-stacking $C_9N_4$. Our results imply possible functionalization for a novel metal-free C-N covalent network with interesting semimetallic properties.

**Introduction**

Graphene[1] discovered in 2004 has fostered a new research filed of two-dimensional (2D) materials, attracting continued research attention. Much effort has been devoted to searching for new 2D materials beyond graphene. Next to C in the periodic table, N appears to be a suitable partner to compose another important set of 2D material with C. The 2D polyaniline with a $C_3N$ stoichiometry has been successfully synthesized by a direct pyrolysis of hexaaminobenzene (HAB) trihydrochloride single crystals very recently.[2] The $C_3N$ monolayer can be considered as N-substituted graphene with uniformly distributed N atoms in an ordered pattern, which possesses an indirect band gap.[3] This C-N compound not only has great potential for applications in nano-electronic but also as a functional unit.[4-6] In comparison with the hole-free honeycomb $C_3N$ monolayer, the holey generated C-N covalent networks [7-12] also draw a great deal of interest lately. Graphitic carbon nitride,[7] g-$C_3N_4$ is extensively studied since 2009, which crystallizes in a porous framework consisting of holes due to large N content. Graphitic $C_3N_4$ has a direct band gap and can be potentially utilized in a wide range of energy and environmental applications, including fuel cells, catalysis, and gas production.[7-9] Likewise, nitrogenated holey grahpene with a $C_2N$ stoichiometry were synthesized by a bottom-up wet-chemical reaction recently.[10] This 2D crystalline $C_2N$ network, with regular holes, has a large electronic band gap of 1.96 eV, making it a promising candidate material for opotoelectronic[10] and gas purification.[11,12] In addition, plenty of C-N materials with different C/N ratios have been theoretically designed via first principle calculations.[13-16]

So far, however, there is rarely any investigation of 2D C-N covalent networks with metallic or better yet NLSM features. The NLSMs as new states of quantum matter have been intensively studied recently.[17-28] Particularly, the C based Mackay-Terrones crystal (MTC),[26] $IrF_4$ calsses,[29] spin-orbit metal $PbTaSe_2$,[30] anti-perovskite $Cu_3PdN$,[24,28] and $Ca_3P_2$ [27] were found to be 3D NLSMs. Attention has also been paid to 2D NLSMs from both experimental research and theoretical design.[18-20, 31-38] The experimental realization of 2D planar $Cu_2Si$ [38] was reported to be a NLSM, in which the nodal lines form two concentric loops centered at Γ point. However, the band touching in $Cu_2Si$ are not rightly lying at the Fermi level and the existence of two nodal loops makes it more challenging for experimental characterization. The first-principle predicted $Hg_3As_2$ [18] and PdS family [19] were only 2D NLSMs in the absence of SOC but became gapped in the presence of SOC.[18, 19]

In this work, we propose a novel covalent network consisting of C and N atoms with a $C_9N_4$ stoichiometry from first principle calculation. Most interestingly, it is found to be a 2D NLSM, in contrast to other C-N compounds [2, 3, 7, 10] as well as COFs [39, 40] which are all insulating. The nodal-line in $C_9N_4$ forms an ideal single loop centered at Γ point without overlapping with any other bands in the momentum space. Mechanistically, this single loop is revealed to be the crossing of two separate sets of bands arising from two sub-lattices, namely the Kagome and Dirac bands. The interpenetration between Kagome and Dirac (honeycomb) lattices proposed recently can lead to different properties, like 2D topological insulator,[41, 42] normal insulator,[43] half metal,[44] Dirac semimetal,[45] and 2D NLSM.[18] In those materials, the sub-lattice sites were occupied by single atoms instead of molecular structures, which are different from our case. The negligible small SOC in $C_9N_4$ is significant for the conservation of nodal-line compared with the annihilation in other 2D NLSM containing heavy metals.[18, 19, 37, 38] Moreover, the large chemical potential difference between C and N also plays a crucial role in the formation of nodal-line in $C_9N_4$, as further confirmed by charge transfer and tight-binding (TB) analysis. Interestingly, there is a new structure of nodal lines formed, i.e. "nodal-cylinder", in the whole Brillouin zone (BZ) along $k_z$ direction for AA-stacking $C_9N_4$. Our results open an avenue for the design of both 2D NLSM and functional covalent organic frameworks.

**Computational methods**

The structure optimization and electronic band calculations are performed within the framework of density functional theory (DFT) using Vienna ab-initio simulation package.[46] The Perdew-Burke-Ernzerhof [47] functional of generalized gradient approximation is employed. The projector-augmented wave [48] method is used to describe the ion-electron interactions. The plane-wave energy cutoff is set to be 500 eV, and the energy convergence criterion in the self-consistency process is $10^{-6}$ eV. A 20 Å length of slab layer is applied along the Z directions to avoid interactions between periodic images. A Γ centered $9 \times 9$ grid of **k** points is used to sample the BZ. The phonon specra is calculated using a 2×2 supercell with Phonopy [49] code. Ab-inito molecular dynamics (AIMD) simulations with NVT ensemble are conducted to investigate the thermal stability of $C_9N_4$ monolayer. The molecular properties are calculated using Gaussian package [50] at B3LYP level.

**Results and Discussion**

As shown in the right panel of Fig. 1, the proposed covalent network has regular periodic nano-holes, in which C and N atoms are represented by black and blue balls respectively. The hexagonal C rings in the network form a Kagome lattice, which is indicated by dashed green line connecting each center of C rings. Whereas the N constitutes form a honeycomb lattice. Overall, C and N atoms share a 2D hexagonal lattice with the same crystal symmetry ($D_6$) as graphene,[1] where the optimized lattice parameter is 9.64 Å. In the left panel of Fig. 1, a C ring joining with six N atoms is shown to form this new type of network. This building block is also the key component of the precursor HAB, which is utilized to synthesize the crystalline $C_3N$ [2] and $C_2N$ [10] compounds. The unit cell of our proposed C-N network, indicated by red lines in Fig. 1, has 18 C atoms and 8 N atoms resulting in a $C_9N_4$ stoichiometry. It is worth noting that $C_3N$, $C_2N$, and $C_9N_4$ have very similar lattice geometries (Fig. S1 in supplemental information (SI)), implying intimate structural relationships between these monolayer carbon nitrides. The $C_2N$ unit cell is $\sqrt{3}\times\sqrt{3}$ times the $C_3N$ unit cell by removing a C ring; while the $C_9N_4$ unit cell is twice as the $C_3N$ unit cell by removing another C ring. According to the atomic chemical environment and symmetry, there are two kinds of C and N atoms in $C_9N_4$. The optimized C-C and C-N bond lengths are similar to those in $C_2N$,[10] which indicates a strong covalency. The perforated covalent network has rich chemically active edge sites with a lone pair, suggests its potential application in chemical sensors.

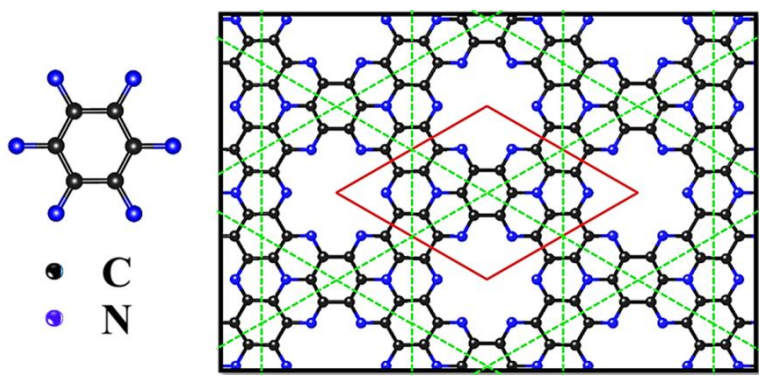

Figure1. The left panel shows the building block for $C_9N_4$ covalent network. The right panel presents the $C_9N_4$ crystal structure, where the red line indicates the unit cell. The C rings form a Kagome lattice connected by green dashed lines.

The uniformly distributed N atoms in $C_9N_4$ network have an atomic size comparable with C and a five-electron valence structure, which makes it a favorable option to realize a strong covalent network. The $C_9N_4$ electronic band structure without SOC is shown in Fig. 2 (a). There are two touching points denoted by D1 and D2 in the proximity of Fermi level (set as zero), which imply a semimetallic nature. This unusual feature is apparently different from previous studied C-N compounds [2, 3, 7, 10] and COFs, [35, 36] which without exception are semiconducting or insulating. To reveal the origin of these two crossing points, the atomic-orbital resolved band structure is plotted in Fig. 2(a), where the sizes of the circles are proportional to the weight of related orbital contributions. It is clear that the two linear crossing bands near the Fermi level stem from the $p_z$ orbitals of C and N atoms. The bands corresponding to the $p_x$ and $p_y$ orbitals lie much deeper below the Fermi level. Furthermore, there are two sets of distinguishable bands, the Kagome and Dirac bands. The former consist of one flat band and two dispersive bands, arising mainly from the $p_z$ orbitals of C; the latter mainly from the $p_z$ orbitals of both C and N, degenerating at **K** points. The two branches of the linear Kagome and Dirac bands cross with each other at points D1 and D2 along Γ-M and Γ-K paths, respectively. To examine whether there are extra crossing points in the whole BZ, we calculated the band structures throughout the first BZ, as shown in Fig. 2(b). Clearly it shows that there are one concave and one convex energy surface crossing with each other around Γ point, forming a single nodal loop in the BZ. Thus, we identify $C_9N_4$ to be a 2D NLSM. The crossing points D1 and D2 are two points in the same nodal ring as seen in Fig. 2(c).

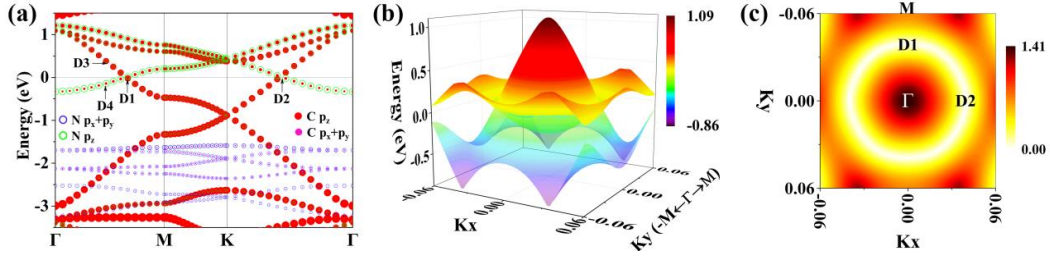

Figure 2. (a) Projected band structure for $C_9N_4$, where the size of dots is proportional to the weight contributed by different orbitals. (b) 3D band structure of two crossing bands in the vicinity of Fermi level around Γ point. (c) 2D projected Fig. of two crossing bands shown in (b), where the color bar indicates the energy difference between conduction and valence bands at each k point.

To compare the states near the two crossing bands, we calculated the charge densities of states at points D3 and D4 as shown in Fig. 3(a), which arise from the Kagome and Dirac bands respectively. The D3 state mainly contributed by the $p_z$ orbitals of C exhibits a Kagome distribution, as indicated by the green hexagons. It results from the C rings connecting each other by an edge hopping path. As shown in Fig. 1, the eight N atoms in the unit cell form a 2×2 honeycomb sub-lattice, however, no band folding feature is observed in the band structure. Nevertheless, the Dirac bands cross at high symmetric K point, implying a bipartite "superlattice". For the D4 state, the Dirac bands can be considered as a super-honeycomb lattice consisting of two super-atoms in the unit cell labeled by the blue triangles displayed in Fig. 3(b). Each of the super-atoms has an identical atomic configuration with $C_3$ rotation symmetry without sharing common atoms. It is noted that the two sub-lattices in the Kagome-honeycomb composite lattice proposed before [18] is represented by single atoms instead of a bipartite "superlattice". Besides, the $p_z$ (on honeycomb sub-lattice) and s (on Kagome sub-lattice) orbitals have opposite parities with respect to the in-plane mirror, which is different from $C_9N_4$ lattice.

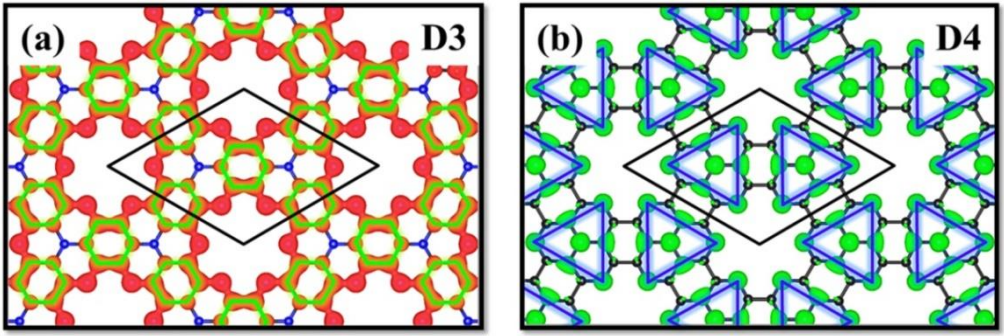

Figure 3. Charge densities for (a) D3 and (b) D4 states decomposed from Kagome and Dirac bands close to Fermi level respectively. The green hexagons and blue triangles form Kagome and honeycomb lattices respectively.

The above calculations are performed without SOC, so it is natural to ask weather this nodal line in $C_9N_4$ is immune to SOC. It is found that the SOC induced gap size is very tiny shown in Fig. S2. Hence, the nodal line in $C_9N_4$ is robust against the intrinsic SOC effect in C and N. We note that the nodal line discovered in the previously reported 2D flat monolayer is mirror-symmetry-protected in the absence of SOC. [18, 38] Meanwhile, the topological index $Z_2=0$, indicating the nodal line in $C_9N_4$ is topologically trivial. To check whether the in-plane mirror symmetry plays a role in forming the nodal loop in $C_9N_4$, we artificially change the planar structure to different buckled structures. As shown in Fig. S3(a-b), two kinds of perturbations without in-plane mirror symmetry are introduced. The corresponding band structures (Fig. S3(d-e)) calculated without SOC clearly show the two touching points along Γ-M and Γ-K directions remain gapless. This evidence shows that the in-plane mirror symmetry has no effect on the protection of nodal loop in planar $C_9N_4$. However, when the sub-lattice symmetry of two super-atoms (Fig. S3 (c)) in the unit cell is broken, implying the chemical potential of two super-atoms differs, the touching point becomes gapped (Fig. S3(f)). The gap size (buckling height) is smaller (larger) in comparison with other 2D nodal line materials containing metal atoms.[19] It is noticed that the degeneracy of Dirac bands at K point is lifted due to the breaking sub-lattice of the bipartite "superlattice", leading to a sizeable gap $E_D$ (Fig. S3(f)). To better understand the effect of chemical potential difference on the formation of nodal loop in $C_9N_4$, a TB analysis is shown below.

Due to the fact that only the $p_z$ orbitals of both C and N atoms contribute to the states close to Fermi level, we can construct a TB model with only the $p_z$ orbitals,

$$\hat{H} = \sum_i \varepsilon_i a_i^+ a_i + \sum_{i \neq j} t_{ij} a_i^+ a_j \quad (1)$$

where $a^+$ and $a$ are electron creation and annihilation operators, $i$, $j$ are the sites of atoms in the unit cell, $t_{ij}$ is the hopping energy between $p_z$ orbitals at site $i$ and $j$, $\varepsilon$ is the onsite energy. We only consider the nearest-neighbor hopping without SOC. As mentioned above, there are two different C and N atoms according to the symmetry and chemical environment. Consequently, four interatomic hopping terms are introduced in the TB model, as shown in Fig. 4(a). The red and gray balls represent N and C atoms respectively, whose size denotes the charge value based on the grid-dependent Bader charge analysis.[51] The calculated results indicate that each N(1) (N(2)) atom gains 1.1 (1.2) electrons, while each C(1) (C(2)) atom loses 0.6 (0.3) electrons. Apparently, the charge transfer happens from C to N due to their electronegativity difference. The TB band structure is calculated to fit the DFT results as shown in Fig. 4(b), where only the five bands comprising three Kagome and two

Dirac bands are plotted. Table 1 lists the corresponding fitting parameters of hopping and onsite energies. The bond lengths between different atoms, for which the hopping integrals are indicated from $t_1$ to $t_4$ (Fig. 4(a)), are 1.39, 1.46, 1.42, and 1.34 Å, respectively. The hopping integrals are about the same since the bond lengths are close to each. However, it is obvious that the onsite energy difference between C and N is rather large. Interestingly, we found that the superposition of the three Kagome bands and the two Dirac bands are mainly attributed to the large onsite energy of N(2) atoms. As illustrated in Fig. 4(c-d), the Dirac point shifted upwards above the flat Kagome band when the onsite energy of N(2) atom is decreased, while the other TB parameters are fixed. Moreover, the width and dispersion of the three Kagome bands are barely affected, and only the Dirac bands are influenced due to the change of onsite-energy of two super-atoms in the unit cell. In particular, when onsite energy of N(2) atoms changes to -1 eV, the linear band crossing disappears as shown from Fig. 4(d). Therefore, the linear crossing is mainly determined by the larger onsite energy of N(2) atoms. Also seen from Fig. 4 (b), the band structures obtained by TB and DFT calculations agree well except the flat band. From DFT results (Fig. 2(a) and Fig. 4(b)), the flat band is pulled down to **K** point and becomes slightly dispersive, which is different from the perfect flat band in TB results. This is because in TB calculation, only nearest-neighbor hopping terms are considered. In fact, such localized flat band in Kagome lattice is known to become dispersive when introducing second nearest neighbor hoppings.[52] Furthermore, in DFT calculations, the linear-crossing bands are contributed by both $p_z$ orbitals of honeycomb and Kagome lattices. Hence, there are hybridizations between the Dirac and Kagome bands. Nevertheless the main feature of linear crossing pertained to the two sets of dispersive bands is captured by the TB model.

To further understand the formation of the intrinsic nodal ring in our 2D covalent system, the properties of molecular orbitals (MOs) for single super atom ($C_9N_4$) and whole unit cell (($C_9N_4$)$_2$) are calculated with Gaussian package. As seen from Fig. S4, the unoccupied MOs from Dirac bands and partially filled MOs from Kagome bands determine the electron filling from the DFT results, i.e., nearly unfilled Dirac band and 2/3 filled Kagome bands. Moreover, the interaction between the wider dispersive Kagome bands and the narrower Dirac bands leads to a crossover giving rise to the degenerate nodal loop. More details are shown in the SI.

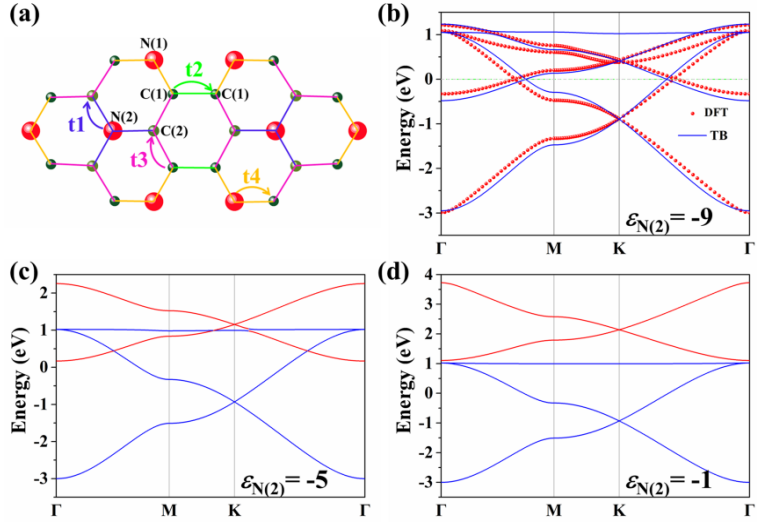

Figure 4. (a) Bader charge analysis for $C_9N_4$ unit cell. The red and gray balls represent N and C atoms. The sizes of balls imply the charge values. (b) Band structure for $C_9N_4$ simulated by TB analysis (blue solid line) compared with DFT results (red dashed line). (c) Band structure for $C_9N_4$ simulated by TB analysis with $\varepsilon_{N(2)} = -5$ eV and (d) $\varepsilon_{N(2)} = -1$ eV, while other parameters are fixed.

Table 1. Values of hopping and onsite energy (units are electronvolts) applied in TB analysis.

|  | $t_1$ | $t_2$ | $t_3$ | $t_4$ |
| --- | --- | --- | --- | --- |
| $p_zp_z$ | 4.0 | 3.8 | 3.9 | 4.1 |
| onsite energy | $\varepsilon_{N(1)}=-5;$ | $\varepsilon_{N(2)}=-9;$ | $\varepsilon_{C(1)}=0.8;$ | $\varepsilon_{C(2)}=0.0$ |

Due to the AA-stacking existed in most COFs, [39] we plot the band structure for AA-stacking $C_9N_4$ shown in Fig. S5. Interestingly, the two linear crossing points shift downwards from $k_z$=0 to $k_z$=0.5 plane. Therefore, the crossing points will form a new structure of nodal lines, i.e. "nodal-cylinder" along $k_z$ direction in the whole BZ, which is different from other simple nodal-line semimetals, [21, 22, 25, 53, 54] nodal-chain metals, [29] nodal-net semimetals, [24, 28] nodal-link semimetals, [17, 55, 56] and nodal-knot semimetals. [57] As shown in Fig. S6, there are two nodal-lines along $k_z$ direction on $k_y$=0 plane, which further confirm the existence of nodal-cylinder. The optimized interlayer length for AA-stacking $C_9N_4$ is 3.45 Å, which is a typical van der Waals length. Other interesting properties related to nodal-cylinder will be studied in future.

Finally, to evaluate the stability of such a new type of C-N network, we next compare the energy of monolayer $C_9N_4$ with pristine graphene, and other experimentally realized 2D C-N materials recently ($C_3N$, $C_2N$, and $g-C_3N_4$). The lower energy for monolayer $C_9N_4$ (Fig. S7) confirms its higher stability. Moreover, the relevant phonon and AIMD computations are performed. There are no soft modes observed in phonon spectrum (Fig. S8), implying the dynamic stability. Further calculation is conducted to study the thermal stability at room temperature (300 K) by performing AIMD simulations. The results (Fig. S8 (b-c)) indicate that there is no structural decomposition observed at 300 K after 10 ps. We also calculated the linear elastic properties from the strain-energy relationship by DFT calculations to investigate the mechanical stability. Due to symmetry, the 2D hexagonal crystal has two independent elastic constants $C_{11}$ and $C_{12}$.[58] The computed values of $C_{11}$ and $C_{12}$ are 208 and 49 N/m respectively, which fulfill the Born stability criteria [59] ($C_{11}>0$, $C_{11}>|C_{12}|$). The other elastic quantities, such as Poisson's ratio and Young's moduli can be evaluated by the following equations,[60, 61]

$$v = \frac{C_{12}}{C_{11}}, \quad E = \frac{C_{11}^2 - C_{12}^2}{C_{11}}. \tag{2}$$

The obtained in-plane Poisson's ratio (Young's moduli) is 0.24 (196 N/m), which is larger (smaller) than those of graphene and $C_3N$.[61] Therefore, our calculations from phonon dispersion, AIMD simulations, and mechanical elastic properties consistently suggest that the $C_9N_4$ network has rather high thermodynamic and mechanical stabilities. All the evidences suggest the synthesis of 2D $C_9N_4$ under ambient conditions is possible.

**Conclusions**

In summary, a new 2D NLSM is predicted using first-principle calculations, which enriches the realm of both COFs and C-N compounds. This novel monolayer with a $C_9N_4$ stoichiometry possesses a nodal-line centered at Γ point in the k-space forming a closed loop. This new feature is rarely found in other C-N compounds and COFs. The nodal-line is mostly contributed by the $p_z$ orbitals of C and N, as further confirmed by the single orbital TB analysis. From the TB results, it is found that the C and N have a large on-site energy difference, particularly, the much larger on-site energy of N(2) atoms plays a significant role in forming such linear crossing at Fermi level. The single nodal loop in $C_9N_4$ is intrinsically immune to the SOC effect due to the super light constituent elements, C and N, which shows an ideal semimetallic properties compared with heavy metal-involved NLSM. Interestingly, a nodal-cylinder forms in the AA-stacking $C_9N_4$ along $k_z$ direction in the whole BZ. The thermodynamic and mechanical stabilities of 2D $C_9N_4$ are guaranteed from our calculations. Our results not only shed light on searching a new kind of 2D C-N material, but also suggest a novel member among COFs with unusual semimetallic properties.

**Conflicts of interest**

There are no conflicts of interest to declare.

**Acknowledgements**

X.N. acknowledges the financial support from the Recruitment Program of Global Young Experts of China and Sichuan one thousand Talents Plan. F.L. acknowledges the support from US-DOE (Grant No. DE-FG02-04ER46148). H.C. acknowledges the financial support from the Graduate School of UESTC. S.Z. is supported by the National Postdoctoral Program for Innovative Talents of China (BX201600091) and the Funding from China Postdoctoral Science Foundation (2017M610858). The computational resources are supported by CHPC at the University of Utah and National SuperComputer Center in Tianjing, China.